# Optimization of collection optics for maximum fidelity in entangled photon sources

Kadir DURAK[1,*]

[1] Department of Electrical and Electronics Engineering, Faculty of Engineering,

Özyeğin University, İstanbul, Türkiye

*Correspondence: kadirac@gmail.com

ORCIDs:

First AUTHOR: https://orcid.org/0000-0002-9333-6309

**Abstract:** In this report the decoherence sources for entangled photons created by spontaneous parametric down conversion phenomenon is studied. The phase and spatial distinguishability of photon pairs from orthogonal crystals reduce the maximum achievable entanglement fidelity. Carefully chosen compensation crystals are used to erase the phase and spatial traces of down conversion origins. Emission angle of photon pairs also leads to optical path difference and resulting in phase distinguishability. A realistic scenario is numerically modelled, where the photon pairs with nonzero emission angle gather a phase difference. These pairs can still be collected and manipulated for practical use but the collection optics adds upon the phase difference. Two commercially available optics for collection; aspheric and achromatic lenses are compared. The numerical simulation results are compared with the experimental results to validate the built model for predicting the maximum achievable entanglement fidelity. The results indicate that the fidelity can be accurately estimated with the presented model by inserting the experimental parameters to it. The study is expected to be very useful for preparation and optimization of entangled photon pair sources in critical phase-matching configuration.

**Key words:** entanglement fidelity, optics, photon pair sources, quantum optics

## 1. Introduction

Quantum entanglement, a peculiar phenomenon of the laws of quantum mechanics, first understood by a series of doubts about the completeness of quantum mechanics after the realization of wavefunction [1,2]. The famous thought experiment for entanglement concept testing and the following experimental observations made the community fascinated by its beauty [3-5]. Although there were still doubts about certain alternative explanation, after closing all three of the known loopholes, quantum mechanical definition of entanglement was accepted by the community [6,7]. Entanglement has emerged as an essential concept and tool for practical quantum technologies including quantum computation, communication and sensing [8-10]. The engineering of entangled photon sources reached to such a high level that they are shown to survive through a rocket explosion, squeezed into a CubeSat and even perform intercontinental quantum key distribution (QKD) [11-13].

Despite the fascinating accomplishments with entangled photon sources there are still open questions and ambiguities on building them. One of the biggest challenges of building an entangled photon source is the trade-off between the brightness and the fidelity [14-16]. The brightness is defined as the number of pairs per unit time per unit input power, pairs/s/mW. Fidelity of a source is a statistical value and can be defined as the strength (or quality) of the quantum correlations between the entangled particles. Entanglement fidelity is shown to be improved by coupling the entangled photons into a single mode fiber (SMF) [17]. However, the brightness is reduced because of the coupling as only photons with certain collection mode are accepted to SMF. The walk-off of the pump and the down converted photons within the crystal also limits the collection to SMF [14]. Fidelity, on the other hand, experimentally maximized until the acceptable value for the application is needed. However, the value is varying even if same non-linear spontaneous parametric down conversion (SPDC) crystal and pump parameters are used [18-21]. The reason for this is that the fidelity is not independent of the collection optics [22]. In this work, the entanglement fidelity with respect to the collection parameters is studied. The emission angle of SPDC and the wavelength of the down converted photons have a correlation. The chromatic dependency of the phase is followed by a numerical model supported by a simulation on the Mathematica software. The aim of the study is to understand the effect of the SPDC parameters including the collection optics and accurately predict the maximum achievable entanglement fidelity.

## 2. Theory

The generation of photon pairs from a second order nonlinear crystal cannot be explained classically. SPDC is the conversion of a high energetic photon interacting with the vacuum state to give birth to two lower energy photons. The term parametric in SPDC indicates that the crystal remains the same after the down conversion process. The daughter photons are called signal and idler. In the SPDC process the energy $\omega_p = \omega_s + \omega_i$ and momentum $\boldsymbol{k}_p = \boldsymbol{k}_s + \boldsymbol{k}_i$ are conserved and the later dictates the phase-matching condition, which governs most of the conversion parameters. Here $p, s$ and $i$ referes to pump, signal and idler, respectively. In this work, a critically phase-matched type-1 Beta Barium Borate (BBO) crystal as the nonlinear medium is considered.

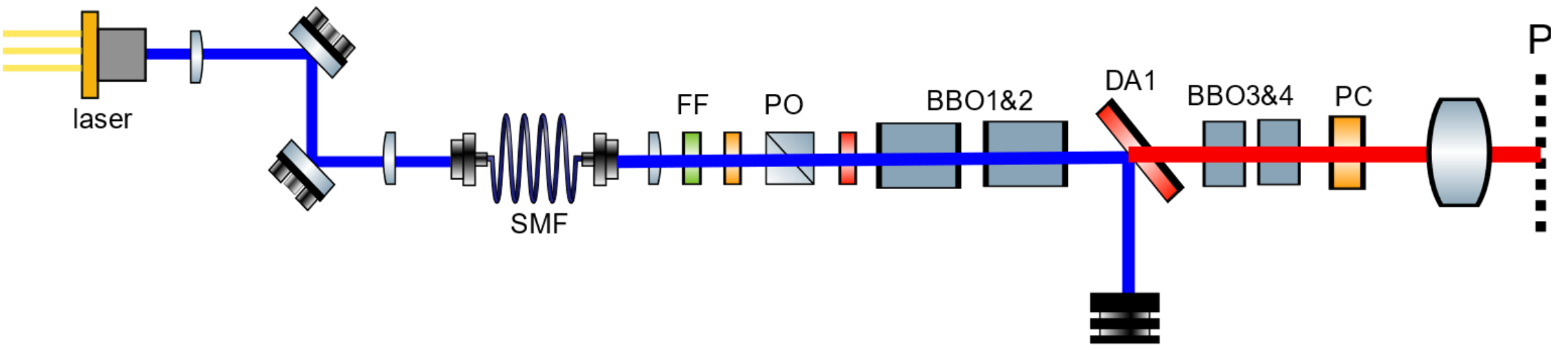

Figure 1 The schematic of the setup used for simulation throughout the paper.

## 2.1. Down conversion

The nonlinear crystal of consideration in this work is BBO and it is a uniaxial crystal such that the horizontally and vertically polarized light experience different refractive indices. BBO can be described with one optical axis and depending on the polarization of light it may experience ordinary $n_o$ and extraordinary $n_e$ refractive indices within the crystal. $n_{o,e}$ are called the principle indices of the medium, and are often determined using an empirical relation known as the Sellmeier formula:

$$n_{o,e}^2(\lambda) = A_{o,e} + \frac{B_{o,e}}{\lambda^2 - C_{o,e}} + E_{o,e}\lambda^2$$

Where the $A_{o,e}, B_{o,e}, C_{o,e}$ and $E_{o,e}$ values can be found in the literature depending on the type of crystal being used [23]. When the light propagates with an angle to the optical axis, the effective refractive index can be expressed as:

$$\frac{1}{n_{eff}^2} = \left(\frac{cos(\theta)}{n_o}\right)^2 + \left(\frac{sin(\theta)}{n_e}\right)^2$$

where $\theta$ is the angle of the propagation direction from the optical axis. The birefringence of the crystals is $n_e - n_o$, and a material is called positive or negative depending on the sign. For angle of propagation with respect to the optical axis $0$ and $\pi/2$, the light experiences a walk-off, which can be expressed as:

$$\rho = \frac{Ln_{eff}^2(\theta)}{2}\left\{\frac{1}{n_o^2} - \frac{1}{n_e^2}\right\} sin(2\theta)$$

where L is the crystal length. The crystal angle, or the optical axis, can now be defined in a way that the propagation direction has a certain angle between the optical axis to satisfy the energy and momentum conservation conditions. BBO is a type-1 crystal such that an extraordinarily polarized pump photon gives birth to two ordinarily polarized daughter photons. In experimental setup the aim is to create the down converted photons with wavelength 780 nm and 842 nm for a pump photon wavelength of 405 nm. This requires the crystal optical axis to be at 28.82º with respect to the crystal facet. This angle is called the crystal cut angle which defines the angle between the propagation direction and the optical axis for a photon propagating normal to the crystal facet.

Figure 1 shows the setup for a typical critically phase matched SPDC configuration. Single mode blue laser light from fiber, polarized at 45º by a half wave plate. Then it passes through the pre-compensator crystal. After that the light gives rise to SPDC while passing through BBO I and II, which are oriented orthogonally with respect to each other. The pump leaves the optical path at this stage and the down converted photons pass through BBO III and IV and the post compensator. The light is analysed at plane P after collected with a collection lens. The photons at plane P usually requires additional steps and care for creating the entangled or Bell's state.

### 2.2. Spatial overlap and phase compensation

The spatial and phase trace of pairs with different polarizations prevents the entanglement due to the distinguishabilities of pairs. There is a set of overlapping and compensation crystals that can be used to achieve the entangled state, which is available in the literature [24].

BBO I and II crystals are orthogonally oriented, which means the down converted pairs from the first crystal will be extraordinarily polarized for the second crystal. Because of this, the pairs from two crystals experience orthogonal walk-off. Therefore, if no counter-measure is taken the horizontal and vertical pairs will be spatially off. This kind of distinguishability between pairs with different polarizations make their polarization deterministic and entanglement cannot be achieved. Therefore, a pair of BBO crystals with half the length of BBO I and II are used to spatially overlap the pairs.

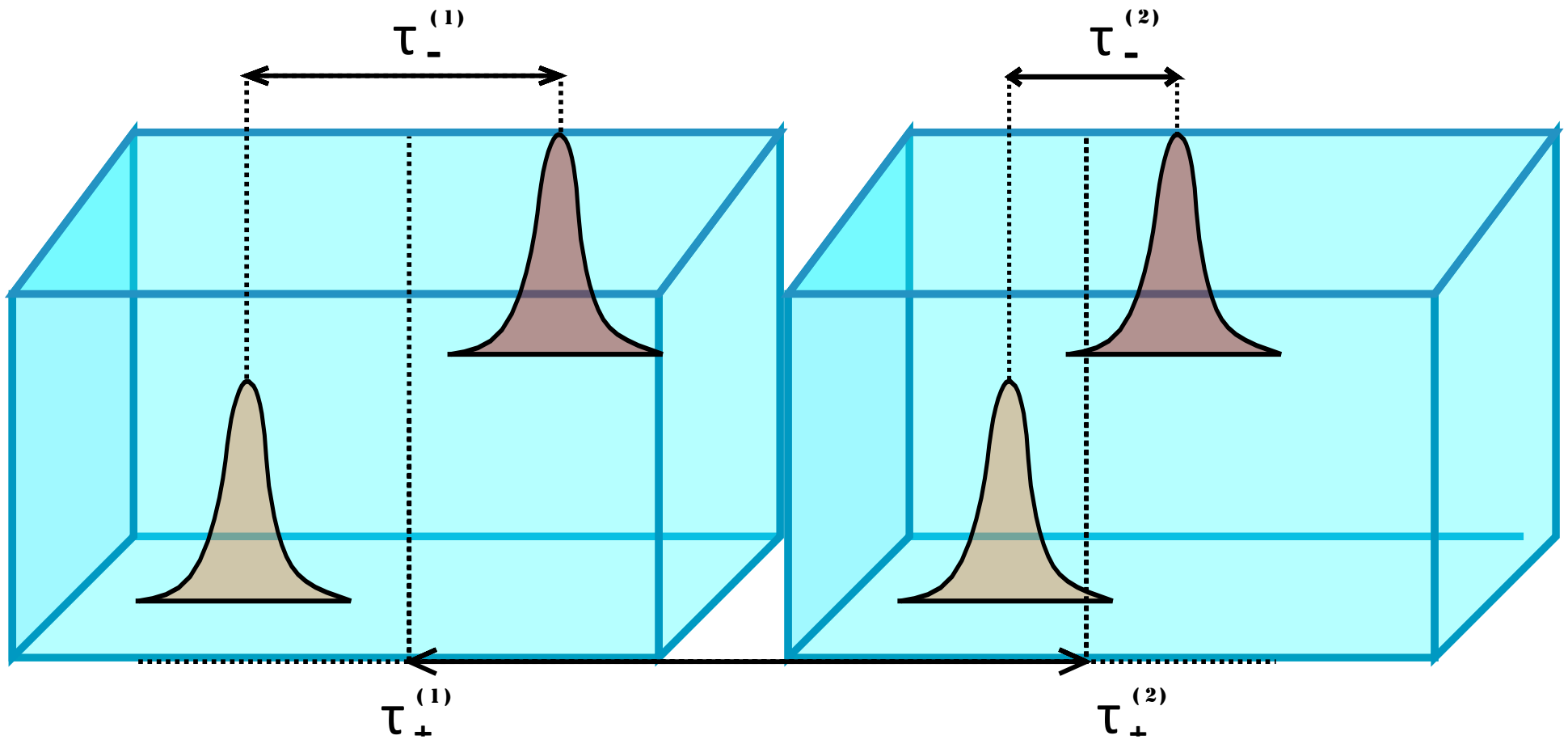

Figure 2 Signal (yellow) and idler (brown) photons from two orthogonally oriented BBO crystals and their relative phase delays are depicted.

There are two types of phase information carried within photon pairs. The first one is due to the birth place of the pairs. The pairs born in the first crystal travel more optical path compared to the ones born in the second crystal and leaving a statistical phase difference, which is shown in Figure 2 as $\tau_+$. By following the measurement times of photons, one can estimate their polarization. In order to erase this phase information an additional $YVO_4$ crystal is used such that the birefringence and the thickness of this crystal compensates for the extra optical path. If the coherence length of the pump is larger than the optical path difference, this phase difference can be neglected.

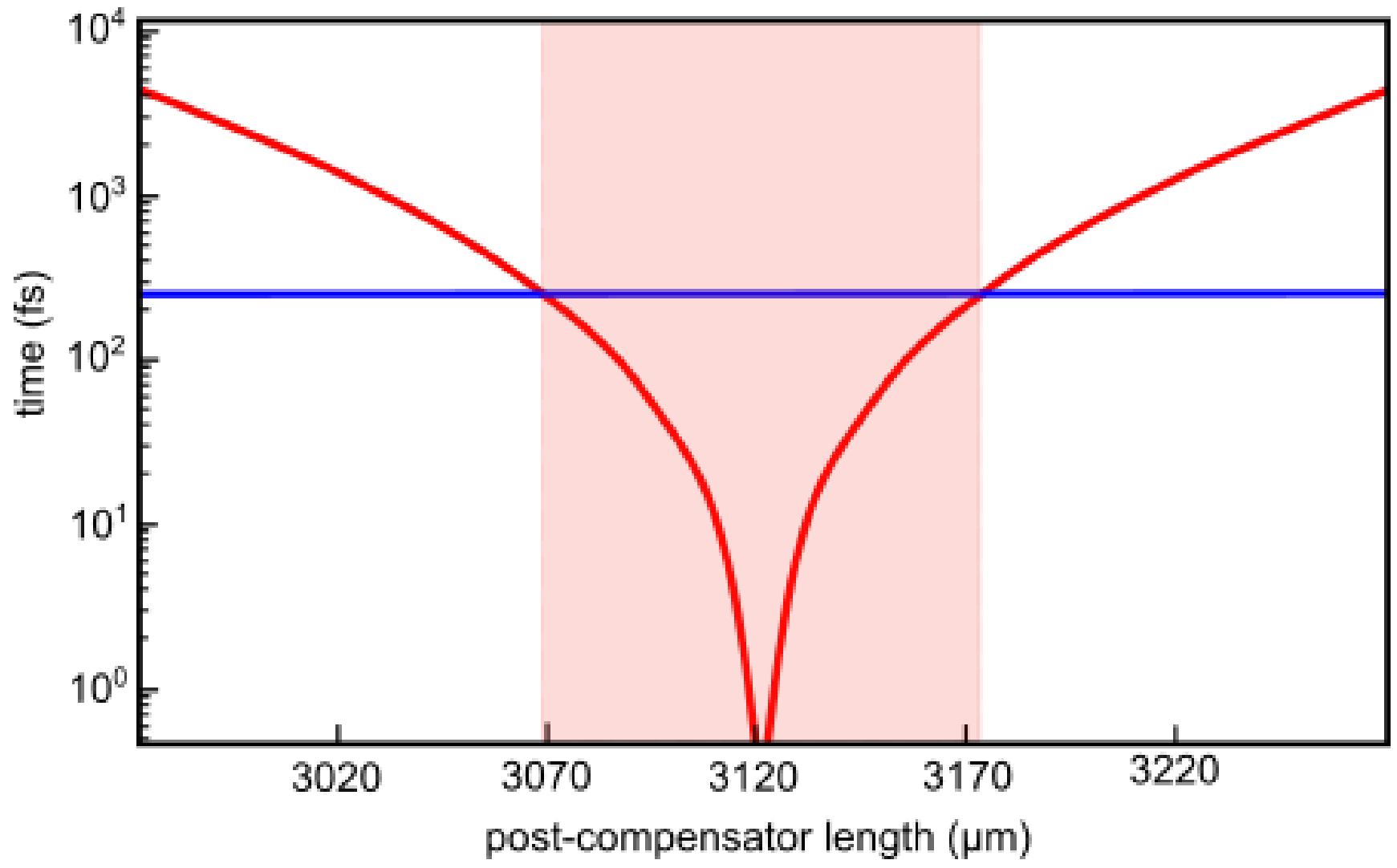

Figure 3 Post compensator crystal length vs the phase difference $\tau_{-}$. The blue line indicates the coherence length of the down converted photons and the pink region shows the allowed error values in crystal length without compromising the indistinguishability.

The second type of phase information is due to the wavelength difference of signal and idler photons for nondegenerate SPDC case. There is a phase difference between signal and idler as the ordinary refractive index depends on the wavelength. However, the phase differences are different for the pairs from the first and the second crystals as the pairs born in the first crystal goes through extra path as extraordinary. This phase difference is shown in Figure 2 as $\tau_{-}$. The difference between signal and idler for different polarization can be traced and therefore entangled state can not be achieved due to the distinguishability of them. A length-wise carefully chosen birefringent crystal, which is called post-compensator, can be used to equalize the phase difference of pairs with different polarizations. This phase compensation is crucial to achieve the Bell state. Figure 3 shows the $\tau_{-}$ phase difference with respect to the error in length of post-compensator $YVO_4$ crystal for 6 mm BBO crystals. The ideal length is 3.12 mm and maximum allowed length tolerance is around ±50 μm for indistinguishability of photon pairs. The coherence time of down converted pairs is calculated for 10 nm signal and idler spectral linewidth values and approximately 230 fs.

### 2.3. Entanglement fidelity

To calculate the distinguishibility of the photon pairs at different phase is to look at their arrival times at a plane. The arrival time information can be converted into distinguishability by the work of R. Rangarajan, in the rest of the study this work will be followed for modelling [25]. The method calculates the joint two-photon complex amplitude (JTPA) of the wavefunction. The density matrix of a quantum state can be written in terms of the arrival times of the signal and idler photons, leading directly to a measure for entanglement fidelity.

In this approach, the two-photon state can be expressed as:

$$|\Psi\rangle = \iint \zeta(t_s, t_i)\hat{a}_s^\dagger(t_s)\hat{a}_i^\dagger(t_i)dt_s\, dt_i|vac\rangle$$

Where $\zeta(t_s, t_i)$ is the JTPA, $|vac\rangle$ is the vaccuum state and $\hat{a}_j^\dagger(t_j)$ the creation operator of a $j$ photon at time $t_j$ with $j$ is either signal or idler. Pump with central frequency $\omega_p$ and bandwidth $\Delta\omega_p$, the detunings of the signal and idler photons can be defined as $\delta_{s,i} = \omega_{s,i} - \frac{\omega_p}{2}$ , and the JTPA can be written as:

$$\zeta(t_s, t_i) = \frac{e^{-i\omega_p \frac{t_s+t_i}{2}}}{2\pi} \iint e^{-i(\delta_s t_s + \delta_i t_i)} e^{-\left(\frac{\delta_s+\delta_i}{\Delta\omega_p}\right)^2} \varphi(\delta_s, \delta_i) d\delta_s d\delta_i$$

The first term outside the integral is the phase determined by the pump photon. The first term in the integral is the additional phase of the two-photon state depending on the time difference of the photon pair and their detuning from the half of the pump frequency, i.e. detuning from the degenerate SPDC wavelength. This additional phase is scaled by the ratio of the detuning of the down-converted field, with respect to the pump linewidth. The last term $\varphi(\delta_s, \delta_i)$, is a real function which calculates the conversion probability pf a pump photon to a photon pair at detunings nondegenerate frequencies $\omega_s$ and $\omega_i$. The last term can explicitly be given as:

$$\varphi(\delta_s, \delta_i) = \left( \frac{sin\left(\frac{\Delta kL}{2}\right)}{\frac{\Delta kL}{2}} \right)^2$$

where L is the interaction length of the SPDC fields (or simply crystal length), and $\Delta k = \left|\boldsymbol{k}_p - \boldsymbol{k}_s - \boldsymbol{k}_i\right|$ is the phase mismatch. The effective density matrix for a polarization-entangled state is obtained by tracing over the time variable as:

$$\rho = \frac{1}{2}(|HH\rangle\langle HH| + |VV\rangle\langle VV| + \mathrm{f}(\Delta t_s, \Delta t_i)|HH\rangle\langle VV| + \mathrm{f}^*(\Delta t_s, \Delta t_i)|VV\rangle\langle HH|)$$

where $\Delta t_j = t_{j,H} - t_{j,V}$ is the time difference between a signal or idler photon with horizontal and vertical polarization. Ray tracing allows to numerically calculate the timing differences for each signal/idler pair from each down conversion crystal with different polarizations. By multiplying these values with the off-diagonal matrix elements one can deduce the fidelity. Therefore, with this model it is possible to use the timing information via the optical path differences of the rays in the simulation to calculate the entanglement fidelity.

## 3. Simulation of SPDC

Mathematica language is used for the standard sequential ray tracing application. Simple refraction and reflection primitives are defined as:

```
reflect[a_, n_] := Module[{c}, c = (a.n)/(n.n); a - 2 c*n]
refract[a_, n_, u_] :=
```

```
Module[{c, aa, n0}, n0 = Sqrt[n.n]; c = (a.n)/n0;
aa = a.a; (a + n/n0*(Sign[c]*Sqrt[c^2 + aa (u^2 - 1)] - c))/u]
```

The rest is defining surfaces, surface normal vectors and refractive indices, which can be calculated using Sellmeier's equation. A ray is defined by its vector components in x and y axes to carry the information of propagation direction; and the point in cartesian coordinates where the optical operation takes place. For this specific application, wavelength and polarization are assigned to individual rays. In order to meet the Poissonian photon statistics of the pump, 100,000 rays are randomly created such that their spatial distribution is Gaussian. The reason of ray number choice is due to the limited computational power of the computer such that the simulation completion time is reasonable. Therefore, the SPDC process is modelled as a stochastic process, where all the properties of the photons are drawn from distributions defined by the underlying physics. The simulation method described in this work is very powerful to investigate the spatial and temporal behaviour of the SPDC photon pairs and even determine the maximum achievable entanglement fidelity.

In the simulation a collimated Gaussian pump with 100 μm waist is used with collinear phase-matching geometry. The distribution of phase matching efficiency with corresponding signal and idler wavelengths for 405 nm pump laser and 28.82$^{o}$ crystal optical axis is shown in Figure 4. The angular distribution is correlated with the wavelength and the collection angle up to 0.36$^{o}$ seems to be sufficient for collecting most of the 780-842 nm pairs. The indication of Figure 4 is that even if collinear geometry is chosen there will be pairs with higher emission angle. This is bad because higher angle photons mean that they have to travel extra path to be collected to a plane. Extra path introduces further phase differences between photons.

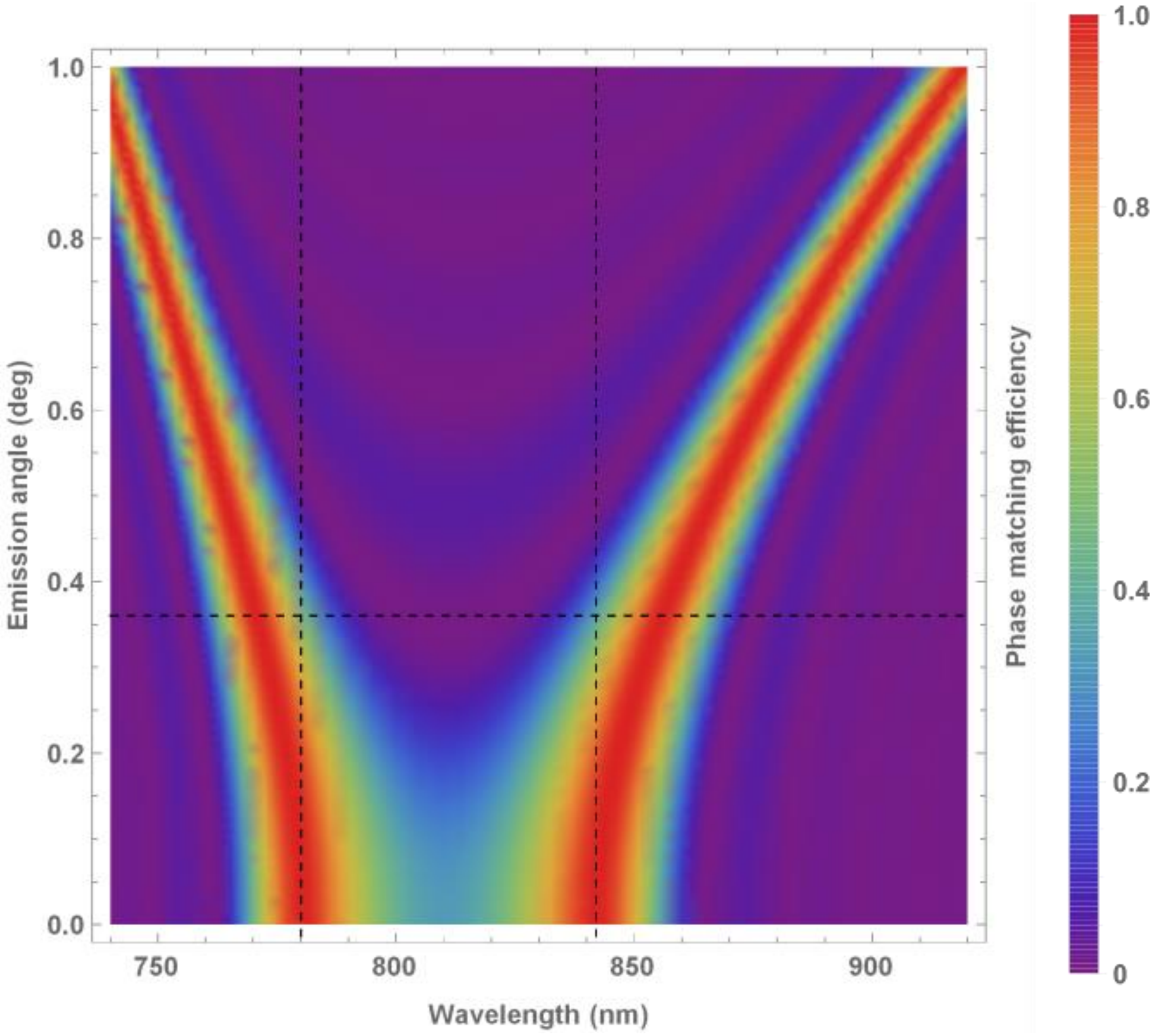

Figure 4 The phase matching efficiency with respect to the emission angle and corresponding wavelength. The left and right branches show the signal and idler photons, respectively.

The additional path for higher SPDC emission angle photons highly depends on the collection optics. Photons travel different paths within different lenses to reach the focus. Figure 5 shows an exemplary focusing of SPDC pairs via an aspheric and achromatic doublet. The upper simulation is performed for Thorlabs aspheric lens A375TM-B and lower one is for Thorlabs achromatic doublet AC-080-16 B-ML. The reason of lens choice is that they are also used in the experimental setup. The aberrations at the focus seems more for the achromatic doublet compared to the aspheric lens.

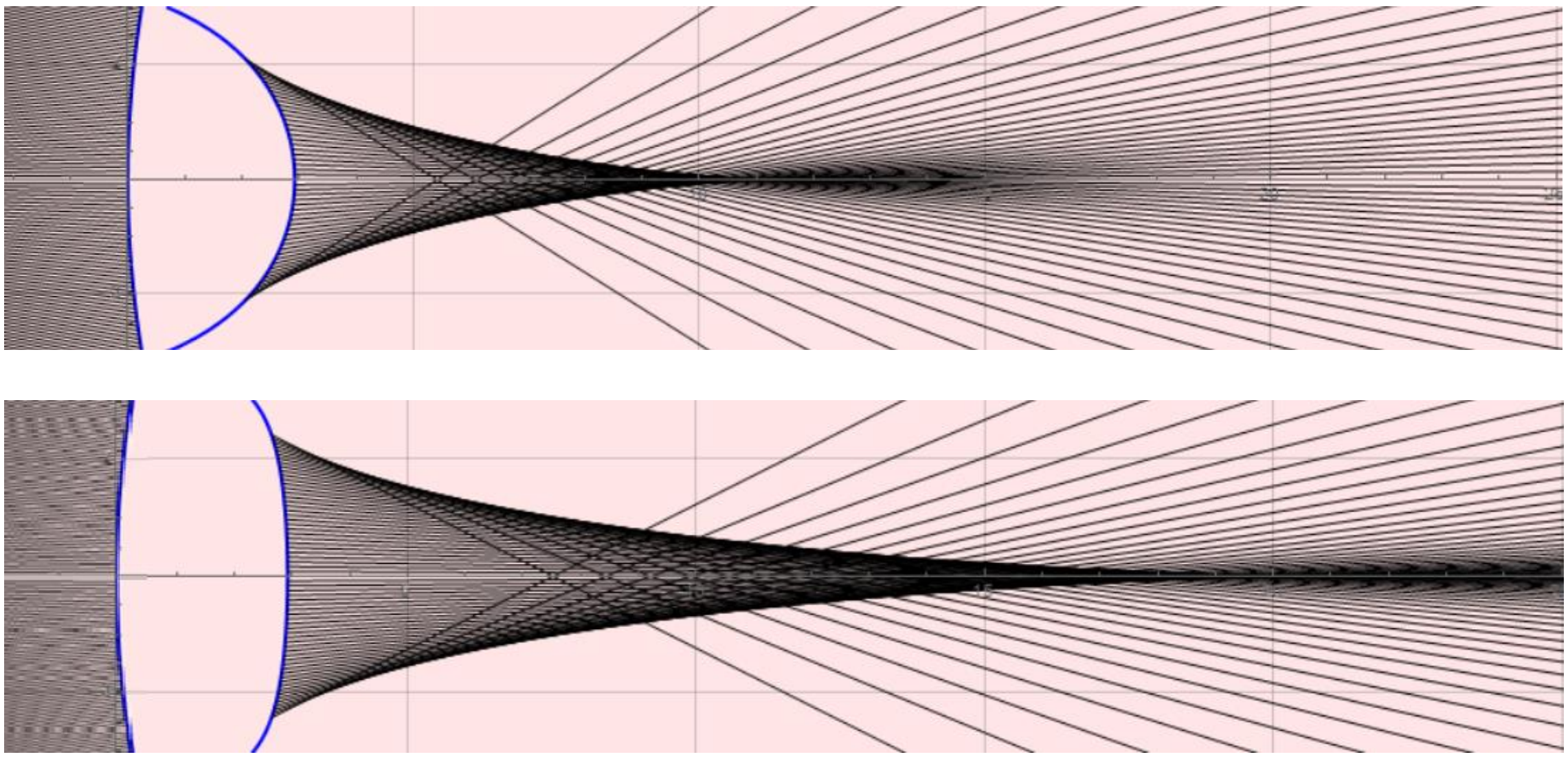

Figure 5 (up) Ray tracing simulations for SPDC collection using aspheric lens, and (below) using an achromatic doublet. For the clarity of the image only 100 rays are used in the Mathematica simulation.

The complete analysis results for 100,000 rays can be seen in Figure 6. The rays are collected at focal plane for both collection optics and the arrival time of individual rays are calculated with respect to the arrival time of the ray at optical axis. The histogram shows the distribution of the $\tau_-$ parameter when the rays are collected with an achromatic doublet (blue) or aspheric lens (red). x axis shows occurrence as the number of rays and y axis shows the parameter $\tau_-$ for the respective delay in femtoseconds. With the arrival timing information of all the rays and using the effective density matrix derived before, it is possible to find the entanglement fidelity. For the achromatic doublet a fidelity of $0.89 \pm 0.11$ and for the aspheric lens $0.98 \pm 0.09$ are found. It should be noted that these values are maximum achievable fidelities as there may be other distinguishability factors in the experimental setup. The reason for the better fidelity with the aspheric lens is that there are smaller number of rays (representing photons pairs) with high arrival time differences.

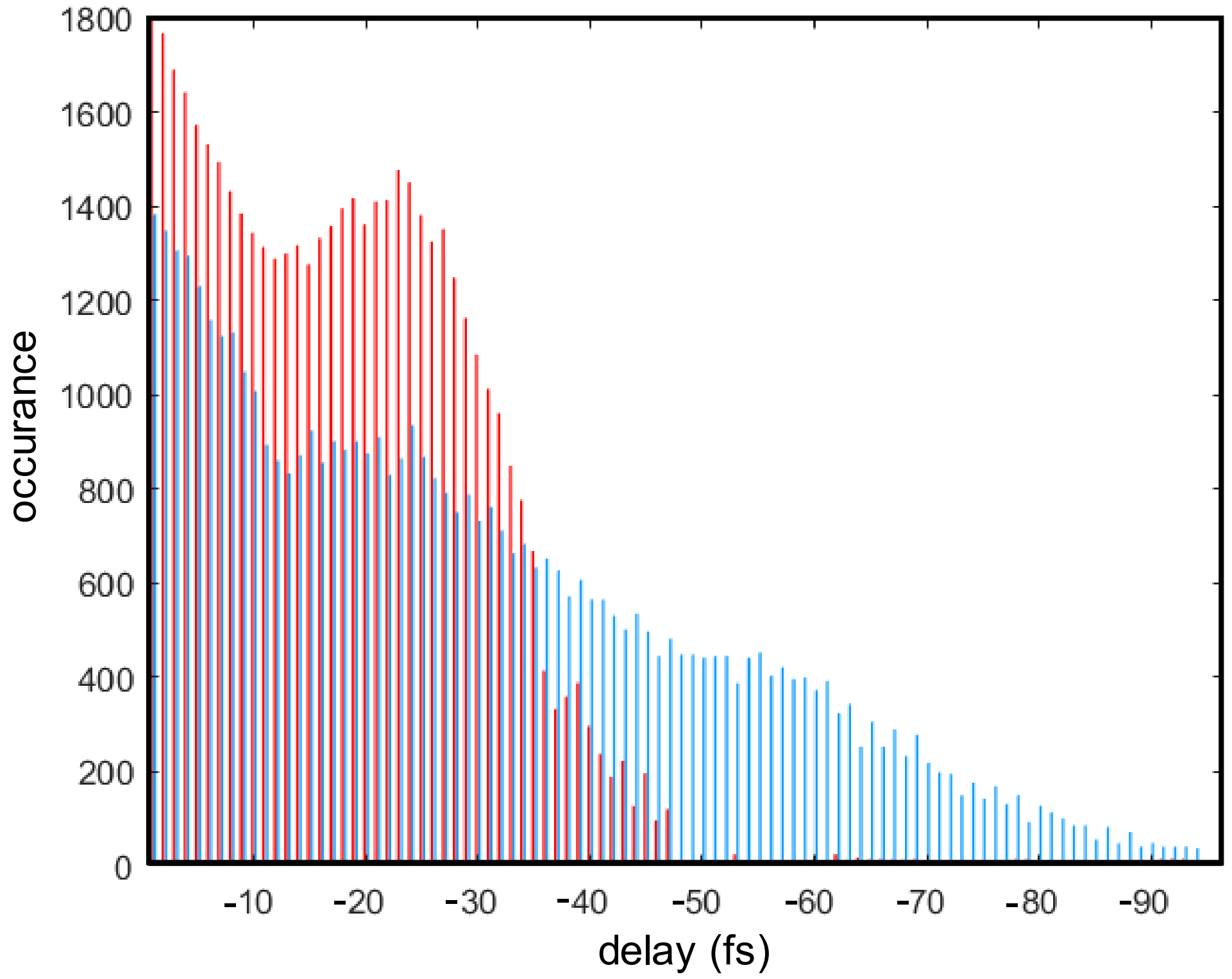

Figure 6 Histogram of the temporal difference of signal and idler $\tau_-$ for aspheric lens (red) and achromatic doublet (blue).

## 4. Experiment

The experimental setup is identical to the one sketched for simulation until the collection optics. In the experimental setup instead of collection optics a dichroic mirror is used to separate the nondegenerate signal and idler photons to different paths for coincidence counting. Coincidences are recorded with varying polarization angle between signal and idler arms to experimentally measure the visibility, which is an indicator of the entanglement fidelity.

Figure 7 shows the schematic of the complete experimental setup. In the detection paths single mode fibers (SMF) are used for collecting highest fidelity photons to get as close as possible to the maximum achievable fidelity. The timing information of the digital pulses from single photon detector D1 and D2 are recorded with high precision to detect coincidences while sweeping the angle between the polarizers on signal and idler paths.

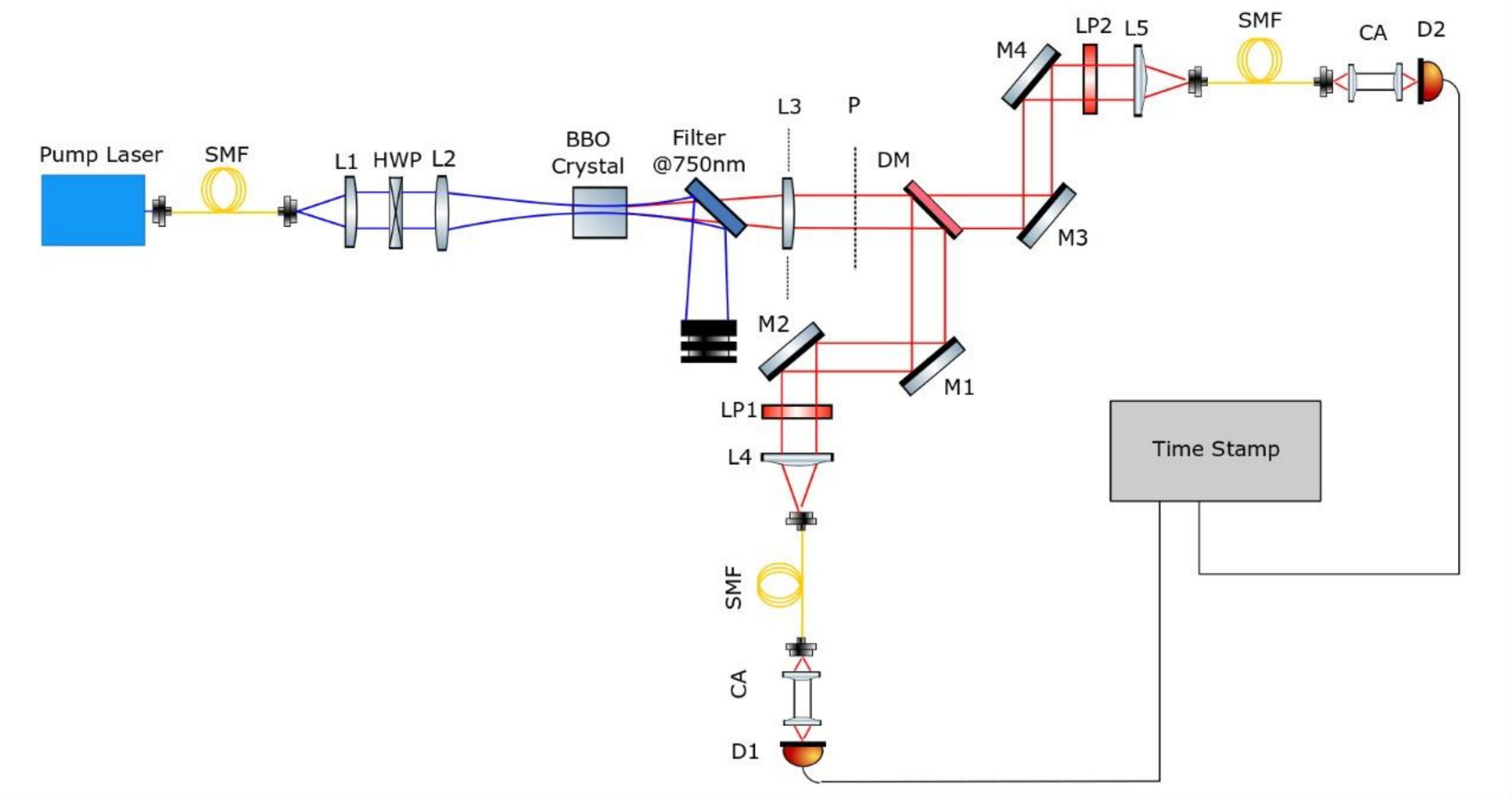

Figure 7 The experimental setup for measurement of the entanglement fidelity.

## 5. Results

The experimental fidelity values are 0.86 ± 0.013 and for the aspheric lens 0.97 ± 0.014. The errors in fidelity values are propagated from the Shot noise of coincidence values. The experimentally observed values are very close to the simulation results for both aspheric lens and achromatic doublet. The aspheric lens seems to be superior to achromatic doublet for both simulation and experimental results. The main reason for this would be the design purposes of these lenses; the achromatic doublet is designed to focus a combination of wavelength to an approximate focus. On the other hand, aspheric lens is designed to focus monochromatic light to focus by keeping the aberrations minimum. The interface and facets of doublet refract the light with different wavelength to different optical paths. Therefore, the optical path difference for rays on different radial distance from the optical axis are different. Aspheric does not seem to suffer from this, as the wavelength difference results in more or less a constant arrival time difference and the radial distance is no concern as the aspheric surface is designed to compensate for the spherical aberrations.

## 6. Discussion

The study presents a method for estimating the maximum achievable entanglement fidelity by numerical simulation of SPDC. The entanglement fidelity can be reached by tracing the rays, which represent SPDC photons, through the arrival time differences of the pairs. A realistic entangled photon source is parametrized within the simulation and commercially available collection optics are tested. The simulation results indicates that a typical aspheric lens is more suitable for collection of entangled photons compared to spherical optics. Experiment is also performed with the exact same parameters with the simulation and the measured results are compatible with the predicted simulation results. This shows that the model, which was present in the literature, combined with the numerical simulation by ray tracing works fine.

The analysis of collection angle relationship with the fidelity can be studied as the collection angle is related to emission angle and wavelength. Therefore, higher collection angle photons are expected to have higher nondegeneracy in terms of wavelength. The presented method can be used to study the mentioned effect.

**Acknowledgement**

This work was supported by Scientific and Technological Research Council of Turkey (TUBITAK) with project number 1109B321700295.